\documentclass[aip,graphicx,amsmath,amssymb,superscriptaddress,reprint]{revtex4-1}

\usepackage[dvipdfmx]{graphicx}
\usepackage{dcolumn}
\usepackage{bm}
\usepackage[colorlinks,linkcolor=blue,citecolor=blue]{hyperref}

\begin{document}

\title{Understanding spin currents from magnon dispersion and polarization: Spin-Seebeck effect and neutron scattering study on Tb$_3$Fe$_5$O$_{12}$}

\author{Y. Kawamoto}
\affiliation{Department of Physics, Tohoku University, Sendai 980-8578, Japan}
\author{T. Kikkawa}
\affiliation{Department of Applied Physics, University of Tokyo, Tokyo 113-8656, Japan}
\affiliation{WPI Advanced Institute for Materials Research, Tohoku University, Sendai, 980-8577, Japan}
\author{M. Kawamata}
\affiliation{Department of Physics, Tohoku University, Sendai 980-8578, Japan}
\author{Y. Umemoto}
\affiliation{Institute for Materials Research, Tohoku University, Sendai 980-8577, Japan}
\author{A. G. Manning}
\affiliation{Australian Nuclear Science and Technology Organisation, Locked Bag 2001, Kirrawee DC, NSW 2232, Australia}
\author{K. C. Rule}
\affiliation{Australian Nuclear Science and Technology Organisation, Locked Bag 2001, Kirrawee DC, NSW 2232, Australia}
\author{K. Ikeuchi}
\affiliation{Institute of Materials Structure Science, High Energy Accelerator Research Organization, Tokai 319-1106, Japan}
\author{K. Kamazawa}
\affiliation{Comprehensive Research Organization for Science and Society, Tokai 300-0811 Japan}
\author{M. Fujita}
\affiliation{Institute for Materials Research, Tohoku University, Sendai 980-8577, Japan}
\author{E. Saitoh}
\affiliation{Department of Applied Physics, University of Tokyo, Tokyo 113-8656, Japan}
\affiliation{Institute for AI and Beyond, The University of Tokyo, Tokyo, 113-8656, Japan}
\affiliation{WPI Advanced Institute for Materials Research, Tohoku University, Sendai, 980-8577, Japan}
\affiliation{Advanced Science Research Center, Japan Atomic Energy Agency, Tokai 319-1195, Japan}
\author{K. Kakurai}
\affiliation{RIKEN Center for Emergent Matter Science, Saitama 351-0198, Japan}
\affiliation{Institute for Multidisciplinary Research for Advanced Materials, Tohoku University, Sendai 980-8577, Japan}
\author{Y. Nambu}
\email[]{nambu@tohoku.ac.jp}
\affiliation{Institute for Materials Research, Tohoku University, Sendai 980-8577, Japan}
\affiliation{Organization for Advanced Studies, Tohoku University, Sendai 980-8577, Japan}
\affiliation{FOREST, Japan Science and Technology Agency, Saitama 332-0012, Japan}

\date{\today}

\begin{abstract}
    Magnon spin currents in the ferrimagnetic garnet Tb$_3$Fe$_5$O$_{12}$ with 4$f$ electrons were examined through the spin-Seebeck effect and neutron scattering measurements.
    The compound shows a magnetic compensation, where the spin-Seebeck signal reverses above and below $T_{\rm comp}=249.5(4)$~K.
    Unpolarized neutron scattering unveils two major magnon branches with finite energy gaps, which are well-explained in the framework of spin-wave theory.
    Their temperature dependencies and the direction of the precession motion of magnetic moments, i.e. magnon polarization, defined using polarized neutrons, explain the reversal at $T_{\rm comp}$ and decay of the spin-Seebeck signals at low temperatures. 
    We illustrate an example that momentum- and energy-resolved microscopic information is a prerequisite to understanding the magnon spin current.
\end{abstract}


\maketitle

Spintronics and magnonics have attracted considerable attention and have gained rapid developments recently.
Creation, annihilation, and control of spin currents--angular momenta flows in the spin degree of freedom--have been central subjects toward future applications.
It can usually be generated electromagnetically~\cite{Silsbee1979,Mizukami2002,Tserkovnyak2002,Kato2004,Wunderlich2005}, optically~\cite{Prins1995}, and thermally~\cite{Uchida2008,Slachter2010,Cornelissen2015}, and the thermal creation via the spin-Seeback effect (SSE)~\cite{Kikkawa2023} has become common recently due to its relative simplicity without using electric and magnetic sources.
The spin current in insulators can be carried by the precession motion of the ordered magnetic moments, i.e., the transverse component of quantized magnons~\cite{Xiao2010}.
Spin currents' propagation goes across an entire material spanning the whole momentum ($\vec{Q}$) space, and the detection via SSE measurements essentially sums up all the $Q$ utilizing the voltage through the inverse spin-Hall effect~\cite{Azevedo2005,Saitoh2006,Valenzuela2006,Kimura2007}.
The measured voltage thus gives the macroscopic sum of the induced spin currents, hence only the overall propagation direction can be discriminated.

This fact makes it challenging to interpret the composition of the overall SSE signals and how the spin current propagates inside the material.
To obtain higher efficiency of the spin currents and to realize future application and implementation to actual devices, microscopic information on the spatiotemporal aspects, or momentum and energy ($E$) in reciprocal space, will be required in addition to the voltage measurements.
Neutron scattering measurements are an effective probe for such systems as they can detect magnons anywhere in the $\vec{Q}$ and $E$ space. 

One successful example using neutrons can be found in the insulating ferrimagnetic garnet Y$_3$Fe$_5$O$_{12}$ (YIG)~\cite{Nambu2021}.
YIG has two major magnon branches, gapless and gapped modes~\cite{Wojtowicz1964,Plant1977,Plant1983,Cherepanov1993,Princep2017,Xie2017,Shamoto2018}, and the mode-resolved direction of the precession motion of the magnetic moment, i.e., magnon polarization, has been defined using polarized neutrons~\cite{Nambu2020}.
The magnon polarization describes integrated energy times the chiral correlation function~\cite{Xiao2010} that governs the spin-Seebeck signal.
At low temperatures, only the gapless mode is thermally occupied, however at room temperature and above, the gapped mode with the opposite polarization starts to be excited and plays a significant role.
Despite the high Curie temperature ($T_{\rm C}=550$~K), a maximum of the spin-Seebeck voltage near room temperature~\cite{Kikkawa2015} has been interpreted in terms of the competition between the two modes contributing with opposite sign of the magnon polarization~\cite{Barker2016,Nambu2020}.
Polarized neutron scattering from YIG thus demonstrated the importance of ($\vec{Q},E$)-resolved information affecting the thermodynamic and transport properties of the compound.

Among the garnet family of crystal structures, other exciting characteristics of the spin current show up when replacing nonmagnetic yttrium with other magnetic rare-earth elements.  
For instance, Gd$_3$Fe$_5$O$_{12}$ (GdIG) possesses two sign changes in the spin-Seebeck voltage against temperature~\cite{Geprags2016}.
The sign change at higher temperature originates from the magnetic compensation~\cite{Geller1965}, while the lower temperature change does not involve any anomaly in magnetization and is thought to involve similar competition between modes with different magnon polarization.
Dy$_3$Fe$_5$O$_{12}$ (DyIG) also involves a sign change at the compensation, but the spin-Seebeck signal under smaller fields is reduced on cooling instead of showing another sign-change~\cite{Cramer2017}.
Those findings highlight the difference of either the quenched (Gd$^{3+}$) or finite (Dy$^{3+}$) orbital degree of freedom, where crystalline electric field (CEF) excitations in DyIG~\cite{Yamamoto1974,Kang2012} may alter magnon dispersion relations.
The large neutron absorption cross-sections of gadolinium (49,700~barn) and dysprosium (994~barn)~\cite{Sear1992} will make neutron measurements difficult, however.

We here raise an intriguing compound with terbium, for which neutron scattering can be applicable owing to the relatively small neutron absorption cross-section of 23.4~barn~\cite{Sear1992}.
Tb$_3$Fe$_5$O$_{12}$ (TbIG) shows a sign change in spin-Seebeck voltage, which is identical to GdIG~\cite{Geprags2016} and DyIG~\cite{Cramer2017}, and the signals nearly vanish at low temperatures without showing the second change.
Magnon polarizations for major magnon modes are clarified above and below the magnetic compensation temperature.
We show that elucidations of magnon dispersion relations including their temperature variation and magnon polarizations, can provide an intuitive prediction of the spin current as a function of temperature.



Polycrystalline samples of Tb$_3$Fe$_5$O$_{12}$ (TbIG) were synthesized by the solid-state reaction prior to single crystal growth.
The stoichiometric ratio of Tb$_4$O$_7$ and Fe$_2$O$_3$ were homogenized by grinding and reacted at 1150~$^{\circ}$C.
Single crystalline samples of TbIG were then grown using the travelling solvent floating zone method.
The quality of the grown single crystals was confirmed by x-ray diffraction. 

\begin{figure*}[t!]
    \centering
    \includegraphics[width=\linewidth]{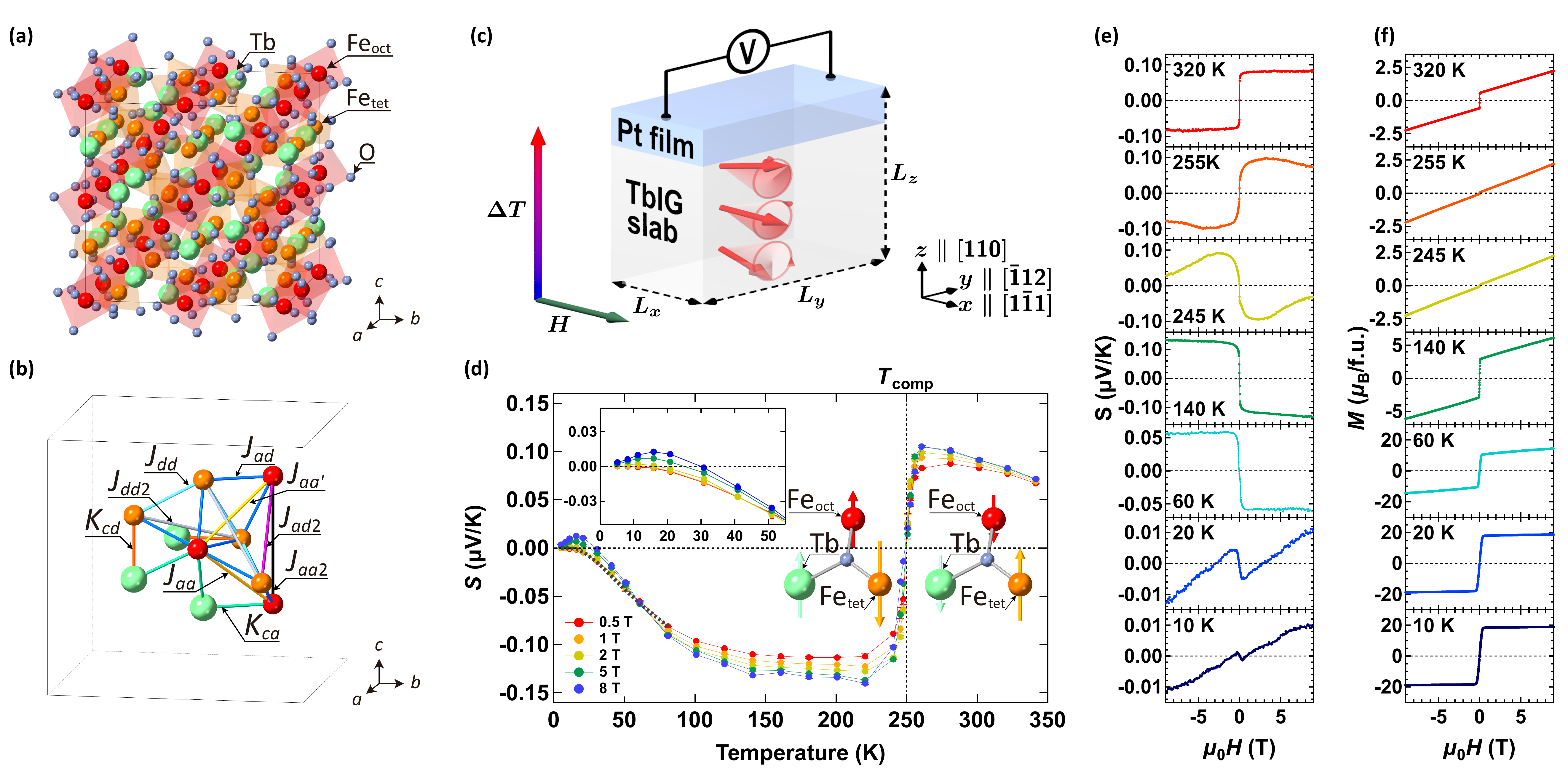}
    \caption{(a) Crystallographic unit cell of Tb$_3$Fe$_5$O$_{12}$ with arrows marking the tetrahedral (Fe$_{\rm tet}$: 24$d$ $(3/8,0,1/4)$), octahedral iron (Fe$_{\rm oct}$: 16$a$ $(0,0,0)$), and terbium (Tb: 24$c$ $(1/8,0,1/4)$) sites. (b) Exchange interactions between Tb, Fe$_{\rm tet}$, and Fe$_{\rm oct}$ used in our spin-wave calculation (see text). (c) A schematic illustration of the longitudinal SSE setup of the Pt/TbIG sample. $\Delta T$, $V$, and $H$ denote the temperature difference along the $z$-direction ($\parallel [110]$), electric voltage between both ends of the Pt layer, and magnetic field applied along the $x$-axis ($\parallel [1\bar{1}1]$), respectively, and $L_x=1.3$, $L_y=4.0$, $L_z=0.6$~mm $+$ 5~nm. (d) Temperature dependence of the transverse thermopower $S$ in the Pt/TbIG sample under various magnetic fields, with the inset enlarging the low-temperature regime. Schematic spin cofigurations above and below $T_{\rm comp}$ are also depicted. Magnetic field dependencies of (e) $S$ and (f) magnetization $M$ measured for representative temperatures. Here, $S$ is defined as $S = (V/L_y)/(\Delta T/L_z)$~\cite{Kikkawa2015,Sola2019,Uchida2016}, enabling the comparison of SSE signals at various temperatures~\cite{Kikkawa2015} and among the garnet family~\cite{Uchida2016}.}
    \label{fig1}
\end{figure*}

Spin-Seebeck effect (SSE) measurements were performed with the longitudinal configuration as illustrated in Fig.~\ref{fig1}(c). 
We used a rectangular-shaped TbIG single crystal, and a 5~nm thick platinum layer was deposited on the crystal using magnetron sputtering.
A magnetic field is applied along the easy-axis direction, $[1 \bar{1} 1]$, to saturate magnetizations, and a temperature gradient is applied along the $[1 1 0]$ direction.
Magnon spin current propagates along the temperature gradient, and angular momentum is transferred to free electrons in nonmagnetic platinum via the interfacial exchange interaction.
Conducting spin current in platinum film is then electrically detected due to the inverse spin-Hall effect.


To examine magnon dispersion relations in TbIG, an inelastic unpolarised neutron scattering experiment was carried out on the chopper spectrometer 4SEASONS~\cite{Kajimoto2011} at J-PARC, Japan.
The instrument enables effective data collection using the multiple incident energies ($E_{\rm i}$) simultaneously.
We employed $E_{\rm i}=128.0$, 49.6, 26.2, 16.1, 10.9, 7.89, and 5.96~meV with energy resolutions of 5.31, 2.75, 1.12, 0.63, 0.41, 0.29, and 0.21~meV at the elastic position, respectively.

To define the magnon polarization, inelastic polarized neutron scattering data were collected on the thermal neutron triple-axis spectrometer Taipan~\cite{Rule2018} at ANSTO, Australia.
Typical components of the instrument are schematically depicted in Fig.~\ref{fig5}(a).
We used a pyrolytic graphite (PG) filter in the outgoing beam to suppress higher-order contaminations and fixed-final energy of $E_{\rm f}=14.87$~meV.

Taipan can accommodate neutron spin-filters with $^3$He-nuclear spins being polarised by the meta-stability optical pumping (MEOP) method~\cite{Colegrove1963,Manning2023}, to polarize the incident neutrons and analyze the scattered neutrons.
Neodymium-based permanent magnets are used to fully saturate the magnetization of TbIG into a single magnetic domain.
The applied horizontal magnetic field (0.38~T) is sufficient to achieve the saturation~\cite{Hur2005} and is set parallel to the scattering wavevector $\vec{Q}$ ($\vec{Q}\parallel \vec{H}\parallel x$) direction.
The neutron polarization was also set by the applied magnetic field in this direction ($P_x$) to detect the chiral term of the cross-sections.
Neutron polarization of the $^3$He cells varies against time, and the spin relaxation lifetime usually exceeds 50~hours in the case of vertical $P_z$ neutron polarization.
In our case, this strong horizontal magnetic field results in stray fields at the position of the spin-filter cells stored in their respective magnetostatic cavities (so-called ``magic boxes''), making the lifetime less than two orders of magnitude smaller than the $P_z$ case.
Despite such short lifetimes, observed neutron-scattering intensities are sufficiently high, hence the neutron polarization correction~\cite{Nambu2023} has not been applied.

The sample used for neutron-scattering experiments on the two different instruments was the same single crystal ($\phi7\times 45$~mm$^2$), and it was oriented on the $(HHL)$ horizontal scattering zone.



The iron-based garnet Tb$_3$Fe$_5$O$_{12}$ (TbIG) is a ferrimagnetic insulator ($T_{\rm C}=560$~K~\cite{Sayetat1984}) with a complex structure, as schematically illustrated in Fig.~\ref{fig1}(a).
The unit cell contains magnetic moments associated with Fe$^{3+}$ ions in tetrahedral (Wyckoff 24$d$ site) and octahedral (16$a$) oxygen cages with opposite directions and moments with Tb$^{3+}$ ions (24$c$) pointing mostly the same direction as the octahedral Fe$^{3+}$ ions.

Figure~\ref{fig1}(d) summarizes the temperature dependence of measured spin-Seebeck voltage with applications of various magnetic fields, which is plotted as a function of the applied field for representative temperatures [Fig.~\ref{fig1}(e)].
With decreasing temperature from $\sim$350~K, the signal stays positive but shows a steep sign-change to negative across 249.5(4)~K.
This involves the magnetic compensation~\cite{Geller1965}, where the sum of Tb, Fe$_{\rm tet}$, and Fe$_{\rm oct}$ moments becomes zero and every moment reverses above and below $T_{\rm comp}=249.5(4)$~K [represented diagrammatically in Fig.~\ref{fig1}(d)].
The observed sign-change in TbIG at $T_{\rm comp}$ is consistent with the findings in GdIG~\cite{Geprags2016} and DyIG~\cite{Cramer2017}.
The magnetic compensation is also evident in the magnetization process [Fig.~\ref{fig1}(f)], where both above and below $T_{\rm comp}$ show a sudden increase with less than 0.5~T yielding ferrimagnetic hysteresis but none is observed at around $T_{\rm comp}$.
Below $T_{\rm comp}$, the voltage stays almost unchanged to 100~K and is steeply suppressed below $\sim$70~K.
For $T<10$~K, the signal takes almost zero at low fields ($\mu_{0}H\le 1$~T), which has stark contrast with the GdIG case showing well-defined voltage even at the same temperature regime~\cite{Geprags2016}.
As enlarged in the left-top inset to Fig.~\ref{fig1}(d), the voltage is slightly enhanced by larger magnetic fields and eventually shows another subtle sign-change.
This is reminiscent of the similar behavior in Dy$_3$Fe$_5$O$_{12}$~\cite{Cramer2017}.
Note that the contribution from the magnetic field ($H$)-linear normal Nernst effect~\cite{Kikkawa2013} in the Pt layer is much smaller than the SSE signal, which was confirmed through Nernst measurements for a Pt/Al$_2$O$_3$/TbIG control sample, with an 8-nm-thick insulating Al$_2$O$_3$ film inserted between the Pt and TbIG layers to block spin currents~\cite{Comment_Nernst}.
The observed $H$-induced sign-change is therefore originating from the SSE.

\begin{figure}[t!]
    \centering
    \includegraphics[width=\linewidth]{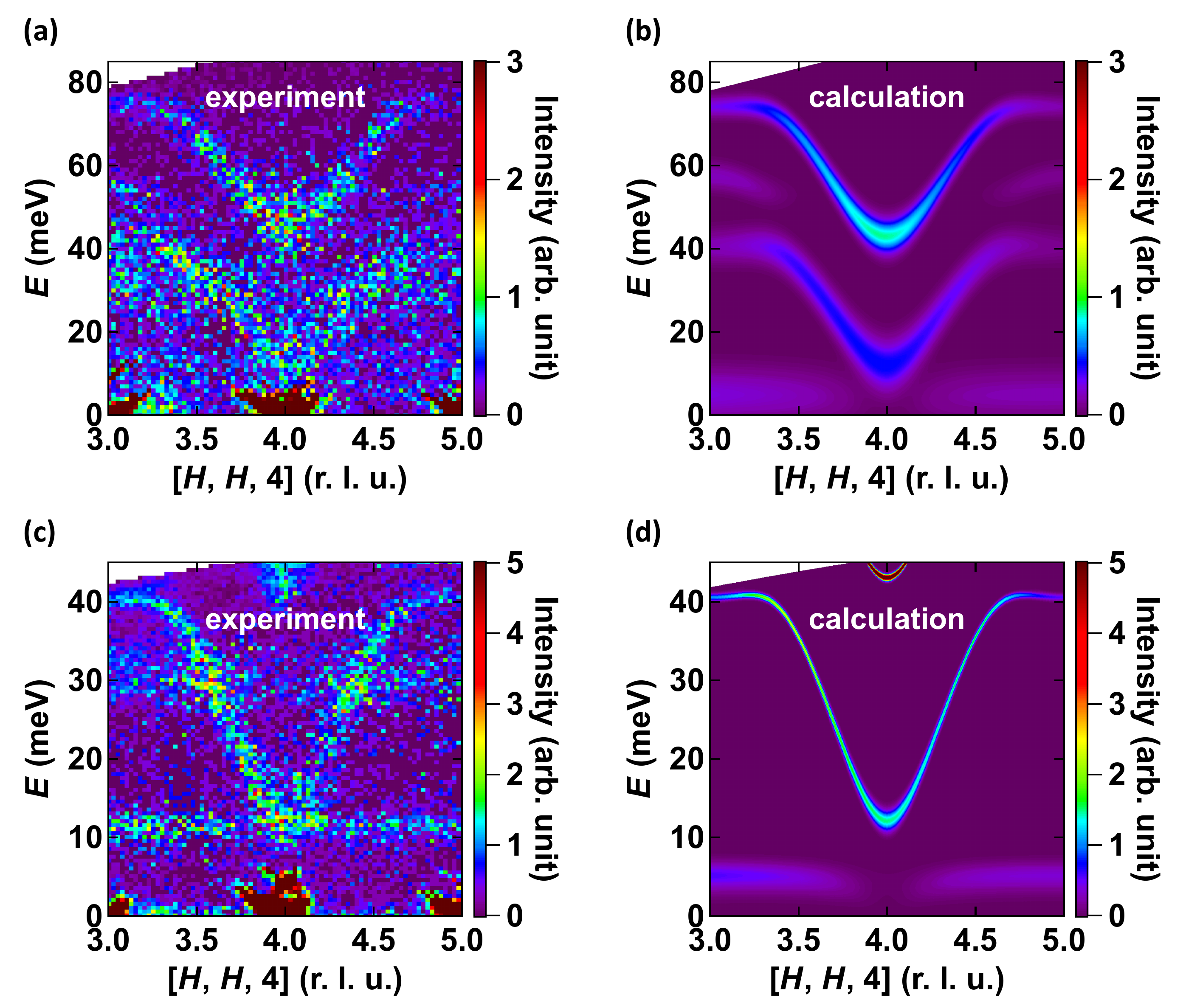}
    \caption{Neutron scattering intensity maps of magnetic excitations in Tb$_3$Fe$_5$O$_{12}$ for (a) $E_{\rm i}=128.0$~meV and (c) 49.6~meV taken at 5~K. Data are sliced along the $[H,H,4]$-direction in the reciprocal space. (b,d) Resolution-convoluted calculated magnon spectra are shown correspondingly.}
    \label{fig2}
\end{figure}

Inelastic neutron scattering has the advantage of measuring magnon dispersion relations and dynamical susceptibility across large areas of reciprocal space.
We performed an inelastic unpolarized neutron scattering experiment using 4SEASONS. 
To capture whole magnetic excitations interested, data with $E_{\rm i}=128.0$~meV taken at 5~K are plotted in Fig.~\ref{fig2}(a).
Two explicit magnon modes with finite energy gaps are visible.
The multi-$E_{\rm i}$ apparatus of the instrument makes it possible to collect smaller $E_{\rm i}$ with tighter energy resolution simultaneously, and Fig.~\ref{fig2}(c) with $E_{\rm i}=49.6$~meV gives an enlarged image depicting the bottom of high-energy mode and entire low-energy mode.

To evaluate superexchange interactions between magnetic ions, obtained data were fit using the random swarm optimization method~\cite{Kennedy1995} within the linear-spin-wave calculation~\cite{Toth2015}.
The spin Hamiltonian we employed reads,
\begin{align}
    \mathcal{H}=&\sum_{\left(i,j\right)\in \left(a,d\right)} J_{ij} \vec{S}^{\rm Fe}_i\cdot\vec{S}^{\rm Fe}_j\nonumber\\
    &+\sum_{c,\ k\in \left(a,d\right)}K_{ck}\left(\vec{S}^{\rm Tb}_c+\vec{L}^{\rm Tb}_c\right)\cdot\vec{S}^{\rm Fe}_k\nonumber\\
    &+D\sum_{c}\left(\vec{S}^{\rm Tb}_c+\vec{L}^{\rm Tb}_c\right)^2,
\end{align}
where $a$, $d$, and $c$ respectively stand for 16$a$ (Fe$^{3+}$), 24$d$ (Fe$^{3+}$), and 24$c$ (Tb$^{3+}$) sites, and $J$ and $K$ describe the interactions within Fe's and between Fe and Tb [all the pathways drawn in Fig.~\ref{fig1}(b)].
The Hamiltonian considers up to the sixth-neighbor interaction for $J$ and the second-neighbor for $K$.
The interactions $J$ up to the third neighbors are all antiferromagnetic, and the best-fit yields $J_{ad}=6.579(5)$, $J_{dd}=0.688(5)$, $J_{aa}=0.000(3)$, and $J_{aa^{\prime}}=0.998(8)$~meV, where $J_{aa}$ and $J_{aa^{\prime}}$ represent the interactions between 16$a$ sites with separate symmetries~\cite{Man2017}.
They were found to take values close to the interactions in YIG~\cite{Cherepanov1993,Plant1977,Princep2017,Nambu2021}.
The fourth- to sixth-neighbor interactions settle in $J_{ad2}=-0.065(2)$, $J_{dd2}=0.031(5)$, and $J_{aa2}=-0.103(5)$~meV.
The interactions between Tb and Fe appear to be relatively weak and are refined as $K_{cd}=0.495(3)$ and $K_{ca}=-0.035(2)$~meV.

Using the above-listed parameters, spin-wave spectra are calculated with resolution convolution [Figs.~\ref{fig2}(b,d)].
The observed and calculated spectra are in good accordance with high- and low-energy modes.
These two parabolic branches are identical to the gapless and gapped modes in YIG~\cite{Wojtowicz1964,Plant1977,Plant1983,Cherepanov1993,Princep2017,Xie2017,Shamoto2018}, but are modulated due to the presence of effective fields from Tb$^{3+}$ ions.
Rather flat excitations at around 10 and 30~meV are apparent in the observed [Figs.~\ref{fig2}(a,b)], which actually come from CEF excitations, as will be discussed later on.
In calculated spectra, on one hand, there appears a weak mode at around 5~meV that is not seen in Figs.~\ref{fig2}(a,c).
This is owing to the interaction between Tb and Fe, corresponding to the gapless mode in the simulated GdIG spectra~\cite{Geprags2016}.
This mode, however, should not emerge since the 4$f$ electrons in Tb$^{3+}$ are actually split into discrete CEF levels, which are not taken into account in the present study.
Once CEF parameters are decided, CEF levels together with spin-wave excitations will be reproduced as exemplified in Yb$_3$Fe$_5$O$_{12}$~\cite{Pecanha-Antonio2022}.
Instead, this study approximates the magnetic anisotropy of Tb$^{3+}$ as Ising-like $D=-0.271$~meV pointing to the $[111]$ and equivalent directions, which is derived from the spin flop transition in the magnetization process~\cite{Lahoubi2012}.
The value of $D$ does not affect the shape of low- and high-energy modes. 

\begin{figure}[t!]
    \centering
    \includegraphics[width=\linewidth]{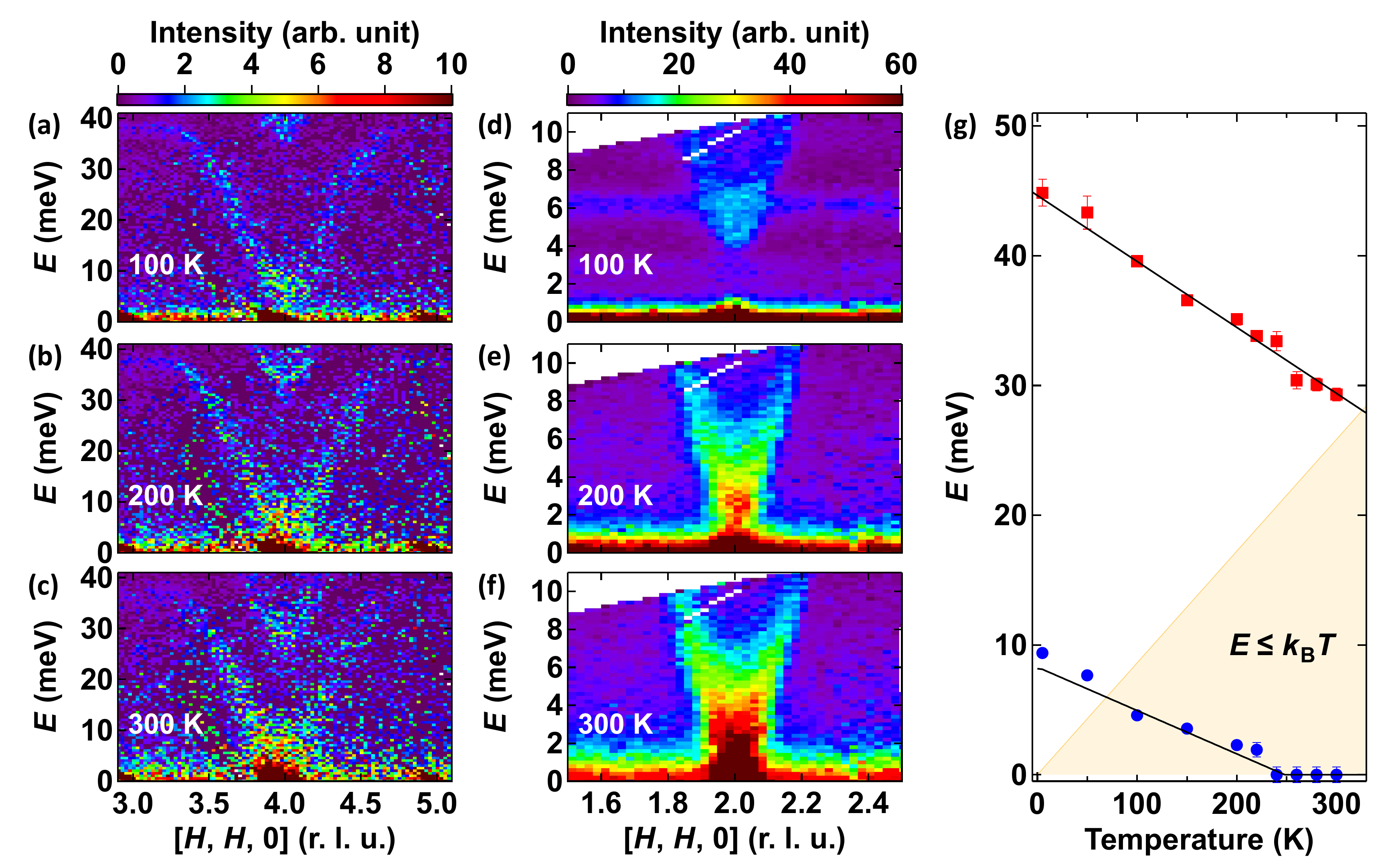}
    \caption{Measured spectra for high-energy (left panels) and low-energy magnon mode (middle panels) taken at (a,d) 100, (b,e) 200, and (c,f) 300~K. (g) Temperature dependence of the estimated gap energies for high-energy (red square) and low-energy magnon mode (blue circle) with linear fits (solid lines). The shaded area marks $E \le k_{\rm B}T$.}
    \label{fig3}
\end{figure}

We also collected temperature dependence measurements of the magnon modes.
Figures~\ref{fig3}(a-f) summarize measured magnon spectra focusing on each mode taken at several temperatures.
Both modes show softening behavior with increasing temperature, and the estimated gap energies as a function of temperature are plotted in Fig.~\ref{fig3}(g).
Magnon modes are thermally occupied below $E=k_{\rm B}T$ [shaded area in Fig.~\ref{fig3}(g)].
The low-energy mode starts to be excited above $\sim$70~K, whereas the thermal excitation of the high-energy mode takes place above $\sim$330~K.

\begin{figure*}[t!]
    \centering
    \includegraphics[width=0.85\linewidth]{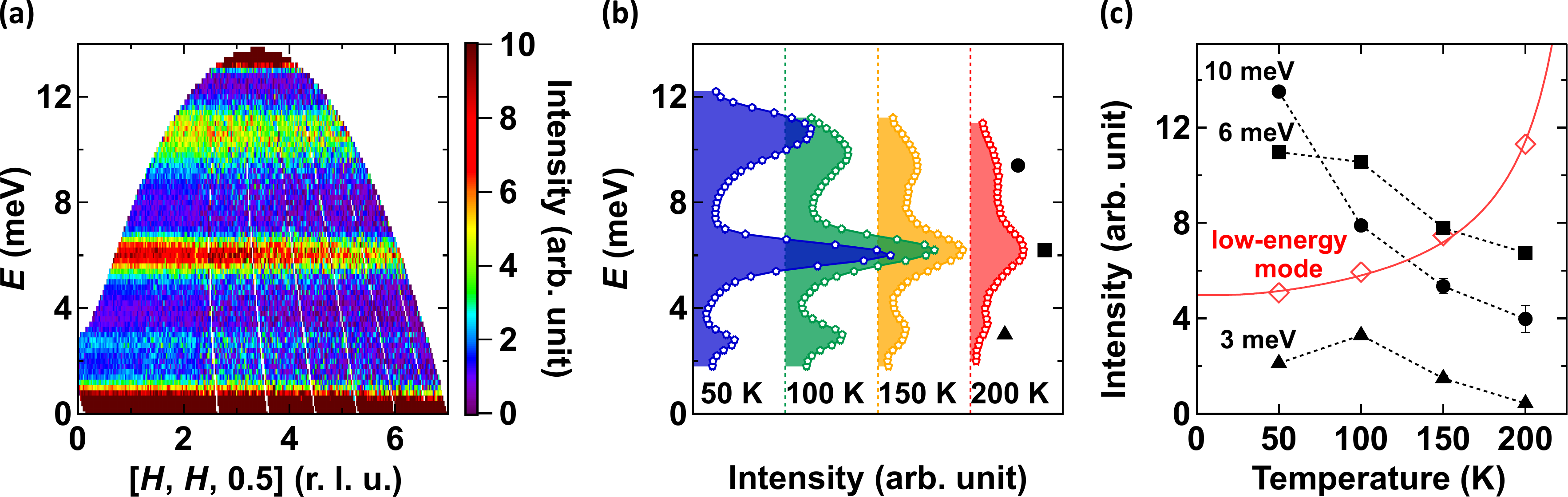}
    \caption{(a) Neutron intensity map with $E_{\rm i}=16.1$~meV taken at 50~K with integrations of $0.35\le [00L]\le 0.65$ and $-0.15\le [K\bar{K}0]\le 0.15$ r.l.u., and (b) derived cuts along the energy direction with the integration of $1.65\le\left|\vec{Q}\right|\le 1.75$ {\AA}$^{-1}$ for representative temperatures. (c) Temperature dependence of the integrated intensities of peaks. The integrated intensity for the low-energy magnon mode obeys the Bose-factor increasing (see text) on warming, whereas flat excitations denoted by circle, square, and triangle do not.}
    \label{fig4}
\end{figure*}

We now turn to the observed flat excitations.
Figure~\ref{fig4}(a) depicts sliced data avoiding any nuclear and magnetic reflections, and flat excitations at around 10, 6, and 3~meV are observed at 50~K, indicating that those excitations appear everywhere in the reciprocal space.
We obtained integrated intensity within $\left|\vec{Q}\right|=1.60\pm 0.05$~{\AA}$^{-1}$ using powder-averaged data, and the energy-spectra are presented in Fig.~\ref{fig4}(b).
Estimated integrated intensities of peaks against temperature are then plotted in Fig.~\ref{fig4}(c).
Excitations from magnons and the CEF can be distinguished by the temperature dependence.
Scattering intensity from magnons when the moment size is irrespective of temperature, increases according to the Bose factor, $\propto\left(\exp(E_{\rm gap}/(k_{\rm B}T))-1\right)^{-1}$ with $E_{\rm gap}$ being the gap energy, and the low-energy mode indeed obeys this law [open red rhomboid in Fig.~\ref{fig4}(c)].
Scattering intensity from the CEF excitations, on the other hand, does not obey that, and instead is dependent on the thermal occupation of both initial and final states during the scattering process.
There is no increase following the Bose factor for the flat excitations [Fig.~\ref{fig4}(c)], and this validates that the ingredients of the flat excitations are the CEF excitations.
Determining the CEF parameters in TbIG is beyond the scope of the present study.


The instrument Taipan [Fig.~\ref{fig5}(a)] was used to define the magnon polarization of the two modes.
The scattered neutrons are recorded in four channels, $I^{+-}$, $I^{-+}$, $I^{++}$, and $I^{--}$, where $I^{io}$ is the intensity of $i$ incoming and $o$ outgoing neutrons with either $+$ or $-$ $P_x$ neutron polarization.
The relation between the scattered neutrons and cross-sections when assuming the perfect performance of all polarization devices are summarized by the following formulae~\cite{Chatterji2006},
\begin{align}
	&I^{\pm\pm}\propto N,\\
	&I^{\pm\mp}\propto M_y + M_z \pm M_{\rm ch},
\end{align}
where cross-sections from isotope-incoherent and nuclear-spin incoherent scattering are omitted for simplicity.
Using the observed four channels, the nuclear ($N$), magnetic ($M=M_{y}+M_{z}$), and chiral ($M_{\rm ch}$) spectra can then be extracted via the combinations,
\begin{align}
    &N = \langle N_Q N_Q^{\dagger}\rangle_{\omega} = \frac{1}{2}(I^{++} + I^{--}),\\
    &M = \langle M_{Qy}M_{Qy}^{\dagger}\rangle_{\omega} + \langle M_{Qz}M_{Qz}^{\dagger}\rangle_{\omega} = \frac{1}{2}(I^{+-} + I^{-+}),\\
    &M_{\rm ch} = i(\langle M_{Qy}M_{Qz}^{\dagger}\rangle_{\omega}-\langle M_{Qz}M_{Qy}^{\dagger}\rangle_{\omega}) = \frac{1}{2}(I^{+-} - I^{-+}),
\end{align}
where $\langle\cdots\rangle_{\omega}$ describes the spatiotemporal Fourier transforms of correlation functions.
$M_{\rm ch}$ consisting of the difference between the spin-flip channles of $I^{+-}$ and $I^{-+}$, describes the chiral correlation function within the plane perpendicular to $\vec{Q}$, and directly relates to the magnon polarization~\cite{Nambu2020}.

At 310~K ($>T_{\rm comp}$), we observed $M_{\rm ch}$ for both modes by making the gap energy of the high-energy mode smaller [Fig.~\ref{fig5}(b)].
The red and blue curves in Fig.~\ref{fig5}(b,c) are deduced chiral terms in the spin-wave spectra, and closed squares give measured $(\vec{Q},E)$-points for the magnon polarization.
Given that observations of the high-energy mode were made in the first attempt in the experiments, the contrast between $I^{+-}$ and $I^{-+}$ is visible but not so pronounced [upper panel in Fig.~\ref{fig5}(d)] reflecting several tens of minutes of lifetime and the ``dark angle'' caused by the magnetic polepiece assembly physically blocking the neutron flight path at particular values of $\vec{Q}$.
Observations of the low-energy mode are from the latest attempt with better stray field shielding, and a few hours of lifetime and fewer effects from the dark angle enable more apparent differences between the spin-flip channels [lower panel in Fig.~\ref{fig5}(d)].
The high- and low-energy modes have magnon polarizations that are opposite in signs from each other as expected from the YIG case~\cite{Nambu2020}.

With cooled down to 160~K ($<T_{\rm comp}$), we measured the low-energy mode with opposite polarization compared to 310~K by moving to the $(2,2,0)$ Brillouin zone to avoid the effects from the dark angle [Fig.~\ref{fig5}(e)].
This indicates that the magnon polarization reverses across $T_{\rm comp}$ accompanied by the reversal of the moment directions and can well be accounted for by the sign change at $T_{\rm comp}$ in the spin-Seebeck voltage [Fig.~\ref{fig1}(d)].
The reason for measuring at this temperature is the collinear magnetic structure, whereas Tb$^{3+}$ moments start to obtain noncollinearity and form an umbrella-type structure below 160~K~\cite{Sayetat1984}.
The gap energy of the high-energy mode at this temperature exceeds 35~meV [Fig.~\ref{fig3}(g)], where the instrument cannot reach such a high energy.

\begin{figure}[t!]
    \centering
    \includegraphics[width=\linewidth]{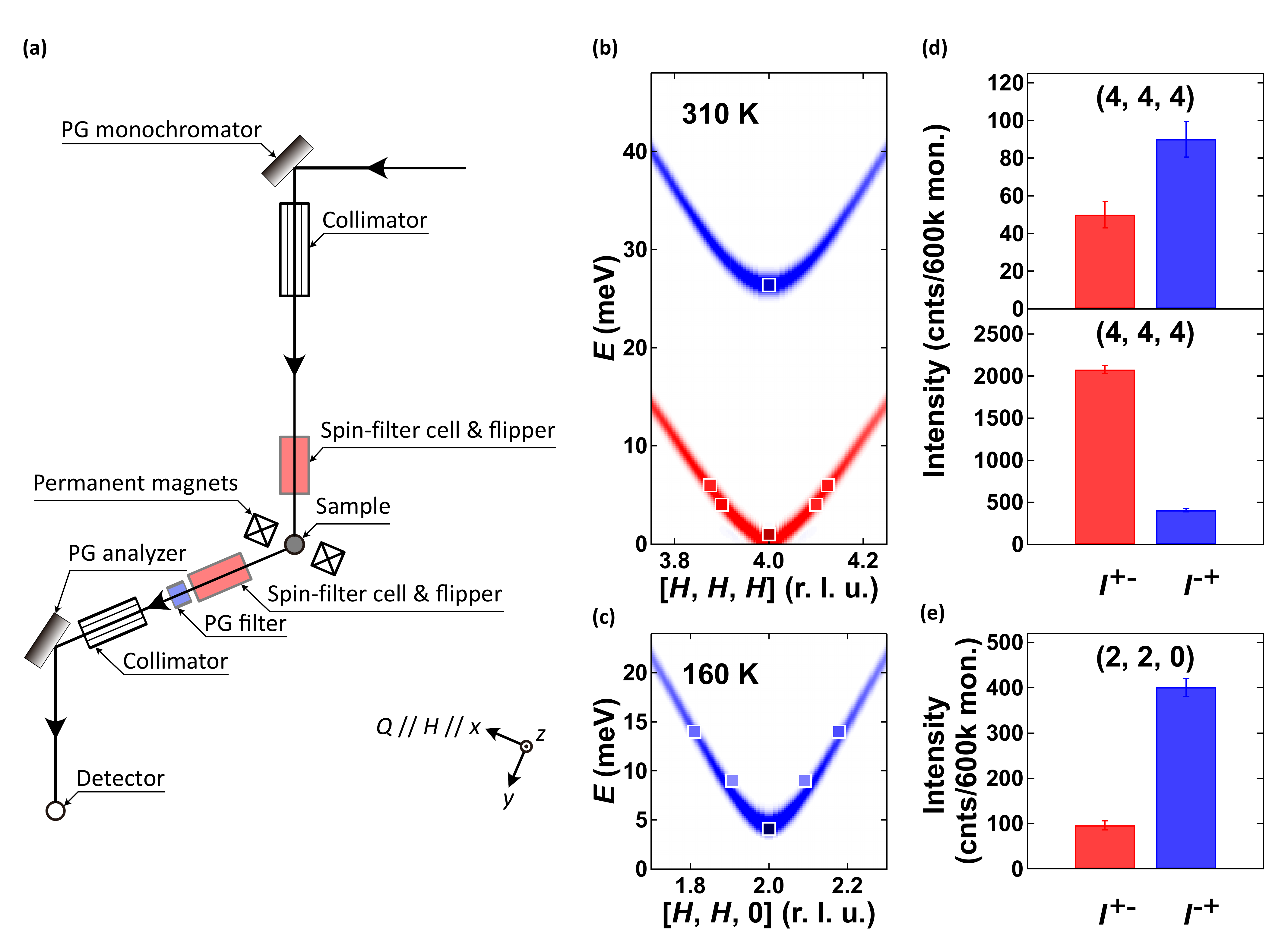}
    \caption{(a) Illustration of the Taipan instrument with bold black arrows denoting the neutron path. Measured $(\vec{Q},E)$ points (closed square) taken at (b) 310 and (c) 160~K together with deduced spin-wave excitations, where the color of each square stands for $(I^{+-}-I^{-+})/(I^{+-}+I^{-+})$. Observed intensity for the spin-flip $I^{+-}$ and $I^{-+}$ channels taken at (d) 310 and (e) 160~K.}
    \label{fig5}
\end{figure}

We also confirmed that the CEF excitations at 3 and 6~meV [Fig.~\ref{fig4}(c)] give no difference in the spin-flip channel, giving confidence that only magnon modes can contribute to the spin-Seebeck signals.
The CEF excitation at 10~meV [Fig.~\ref{fig4}(c)] has $I^{+-}/I^{-+}=59(8)/38(6)$ intensities, where the calculated weak mode at around 5~meV [Fig.~\ref{fig2}(d)] is possibly merged due to the presence of CEF levels and contributes to the slight difference.
Again, the low-energy mode can only be thermally excited below room temperature and is thus responsible for the temperature dependence of the signals [Fig.~\ref{fig3}(g)].
Below 70~K, however, the low-energy mode is no longer excited hence the carrier for the spin current is missing, yielding an explanation of the reduction of the signal on cooling.
The data below 80~K is well fit by the exponential decay [dotted curve in Fig.\ref{fig1}(d)], giving the gap energy of 7.32(8)~meV which is consistent with the results in Fig.~\ref{fig3}(g).
This contrasts with the YIG~\cite{Kikkawa2015} and GdIG~\cite{Geprags2016} cases, where well-defined spin-Seebeck signals are still present owing to the presence of the gapless magnon mode.
In TbIG, the competition between two modes takes place above 330~K, with the high-energy mode being excited, having a relation to the observed reduction of the signal [Fig.~\ref{fig1}(d)].
These findings conclude that the temperature evolution of spin current is essentially predictable from magnon dispersion and polarization.
Theories handling the interfacial exchange interaction~\cite{Adachi2011} are currently under development but yet to be well established, while microscopic views of magnon modes give an intuitive understanding of the spin current.
Although determining the CEF parameters and elucidating the subtle low-temperature sign-change with higher fields remain future subjects, the idea is generally applicable to the spin current in insulating ferromagnetic and ferrimagnetic systems.


To summarize, we measured the spin current-induced voltage via SSE, and clarified magnon dispersion relations including their temperature dependence and magnon polarization of Tb$_3$Fe$_5$O$_{12}$.
The observed two magnon modes are well reproduced by the spin-wave calculation using nine sorts of interactions and the anticipated Ising-like anisotropy.
Temperature dependence of the magnon modes and their polarizations give clear explanations of the spin-Seebeck signals, which show the sign-change across $T_{\rm comp}$ and a rather faster decay as $T\rightarrow 0$.
Unlike the GdIG and YIG cases, both magnon branches are gapped in TbIG, giving rise to no sign change in the spin-Seebeck signals and instead the exponential decay on cooling.
Besides that, the flat modes derived from the CEF excitations cannot carry the spin current reflecting zero group velocity.
Magnon spin currents in insulating magnets can generally be understood once the magnon dispersion relations with their temperature variation, and magnon polarization are clarified.
Our findings have thus demonstrated the importance of $(\vec{Q},E)$-resolved information for spintronics and magnonics.

We thank J. Barker, G. E. W. Bauer, Y. Ikeda, M. Mori, J. Nasu, Y. Onose, and T. J. Sato for valuable discussions.
This work was supported by the JSPS (Nos.~21H03732, 22H05145, 19K21031, 19H05600, 22K18686, 22H05114), ERATO ``Spin Quantum Rectification Project'' (No.~JPMJER1402), CREST (Nos.~JPMJCR20C1, JPMJCR20T2), and FOREST (No.~JPMJFR202V) from JST, Institute for AI and Beyond of the University of Tokyo, and Collaborative Research Center on Energy Materials at Institute for Materials Research, Tohoku University.
4SEASONS experiment was carried out through the proposal No.~2018B0207.
Work at ANSTO was supported by the Graduate Program in Spintronics at Tohoku University.

\medskip

\noindent
\textbf{AUTHOR DECLARATIONS}

\noindent
\textbf{Conflict of Interest}

The authors have no conflicts to disclose.

\medskip

\noindent
\textbf{Author Contributions}

Y.~N., T.~K., and E.~S. conceived the idea of the project.
T.~K. and E.~S. performed the spin-Seebeck effect measurements.
Y.~K., M.~K., K.~Kakurai, and Y.~N. performed neutron scattering experiments with support from A.~G.~M., K.~C.~R., K.~I., and K.~Kamazawa.
Y.~K. and Y.~N. analyzed the experimental data and Y.~K., Y.~U. and Y.~N. prepared figures.
Y.~K. and Y.~N. wrote the paper, and all authors discussed the results in the manuscript.

\medskip

\noindent
\textbf{DATA AVAILABILITY}

The data that support the findings of this study are available from the corresponding author upon reasonable request.


\begin{thebibliography}{}
\bibitem{Comment_Nernst} The $H$-slope for the normal Nernst effect in Pt was evaluated to be $\sim 0.12~\textrm{nV/KT}$ at $T=10$~K, which is one order of magnitude smaller than that for the observed SSE at the same temperature [$S/\mu_0H \sim 1.4$~nV/KT for 0.5 $\leq \mu_0H \leq$ 9~T in Fig.~\ref{fig1}(e)].
\end{thebibliography}
\bibliographystyle{}

\end{document}